\documentclass[authoryear,preprint,12pt,english,review]{elsarticle}
\usepackage{amsmath}
\usepackage{amssymb}
\usepackage{graphicx}
\usepackage{lineno}

\usepackage[T1]{fontenc}
\usepackage[utf8]{inputenc}

\journal{JASTP}

\begin{document}
\begin{frontmatter}



\title{Dynamics of vertical ionospheric inhomogeneities over Irkutsk during 06:00-06:20UT 11/03/2011 caused by the Tohoku earthquake}


\author{O.I. Berngardt}
\ead{berng@iszf.irk.ru}

\author{G.V. Kotovich}
\author{S.Ya. Mikhailov}
\author{A.V. Podlesny}
\address{Institute of Solar-Terrestrial physics, P.O.Box 291, Irkutsk, 664033, Russia}

\begin{abstract}

According to the data of quasi-vertical sounding at Usolie-Tory path (midpoint - 52.3N, 103E, 120km distance),  the dynamics of vertical structure of the midscale inhomogeneities of the plasma frequency was reconstructed. 

These irregularities are caused by Tohoku earthquake 11/03/2011 (38.3N, 142.4E).  Observed horizontal velocities of propagation in the ionosphere and  the spatial scales of the wave-like disturbance modes,  generated by earthquakes, are calculated.

As a result of numerical simulation involving the data from TALAYA seismic station (TLY, 51.7N, 103.7E) and comparison with the experiment it is shown that the vertical ionospheric irregularities of 5-40km range observed in first 20 minutes after the beginning of observations of effects (from 06:00 to 06:20 UT) are qualitatively explained by travelling of acoustic shock wave cone ( Mach cone ) from supersonic ground source - seismic wave to the heights up to 200km.

It is shown that the most likely source of the shock wave were Z and E components of the seismic oscillations observed at TLY station. Irregularities observed after 06:20UT were apparently linked with other mechanisms. It is shown that the existing temporary CHIRP ionosonde temporal resolution (60 seconds) and the ionogram inversion technique used are not enough for detailed diagnosis of fine spatial vertical structures of this process.

\end{abstract}
\end{frontmatter}


\section{Introduction}

Tohoku earthquake (11/03/2011 05:46:24 UT) at the moment is considered to be one of the most powerful earthquake in the world since 1900 and one of the most investigated. Ionospheric effects of the Tohoku earthquake are currently the subject of many papers. Above all, it is papers analyzing data from dense networks of GPS-receivers, located on the Asian territory \citep{Akhoondzadeh_2012, Kakinami_et_al_2011, Astafyeva_et_al_2013, Chen_et_al_2011, Hao_et_al_2012, Komjathy_et_al_2012, Liu_2012,Liu_et_al_2011, Occhipinti_et_al_2013, Ogawa_et_al_2012, Ouzounov_et_al_2011, Saito_et_al_2011, Tsugawa_et_al_2011} and data from the SuperDARN radar, located on the  Hokkaido island \citep{Nishitani_et_al_2011, Ogawa_et_al_2012}. These instruments have the best temporal resolution - from several seconds to 2 minutes, enabling to monitor the highly dynamic processes.

Investigation of more dynamic ionospheric effects over the Czech Republic according to the Doppler radars \citep {Chum_et_al_2012} with 1.6-second time resolution showed that the variation of the Doppler drift velocities correlates well with surface seismic waves.

There are many papers devoted to ionospheric precursors to earthquakes \citep{Carter_et_al_2013, Choi_et_al_2012, Heki_2011, Kamogawa_and_Kakinami_2013, Le_et_al_2013, Zolotov_et_al_2012, Zhou_et_al_2013} and to results of modeling of ionospheric effects \citep{Galvan_et_al_2012, Kherani_et_al_2012, Makela_et_al_2011, Matsumura_et_al_2011, Shinagawa_et_al_2013, Maruyama_and_Shinagawa_2014}, based on the data from these instruments.

GPS networks and SuperDARN radars have high temporal and spatial resolution but low height resolution, so the data from these instruments are used to analyze mostly horizontal structure of ionospheric irregularities and its temporal dynamics.

However, these methods can not track the vertical structure of the irregularities in the ionospheric height. At the present time, the most widely-used tools for the study of the vertical structure of ionospheric irregularities are (in addition to 
incoherent scatter radars) ionosondes of different types - both impulse and continuous (CHIRP).

According to the data of ionosondes, which enable high-precision study of the vertical profile of the electron density,   the vertical structures in the plasma frequency profile with the characteristic scale 10-20km occurred in the region of the earthquake. These structures were very dynamic \citep{Liu_and_Sun_2011, Maruyama_et_al_2011}. However, traditionally ionosondes have relatively low (about 15 minutes) temporal resolution, not allowing us to study dynamics of the vertical structure of irregularities.

In this paper, the dynamics of these vertical ionospheric irregularities of 5-40km scale is analyzed based on the quasi-vertical Ionosonde of the Institute of Solar-Terrestrial Physics (ISTP SB RAS) with high temporal resolution (1 minute). Measurements were carried out at a distance of 3400km from the earthquake epicenter, in Irkutsk, Russia (52N, 103E). When analyzing we used data of TALAYA seismic station(51.7N, 103.7E), located near the region of ionospheric observations.

\section{Description of the experiment}

The basic equipment, used for the data analysis, is bistatic CHIRP Ionosonde made by ISTP SB RAS \citep{Podlesny_et_al_2011}, providing sweep speeds up to 1000kHz/sec and providing temporal resolution 60 seconds comparable to the SuperDARN radar network and GPS-receivers.

\begin{figure}
\includegraphics[scale=0.5]{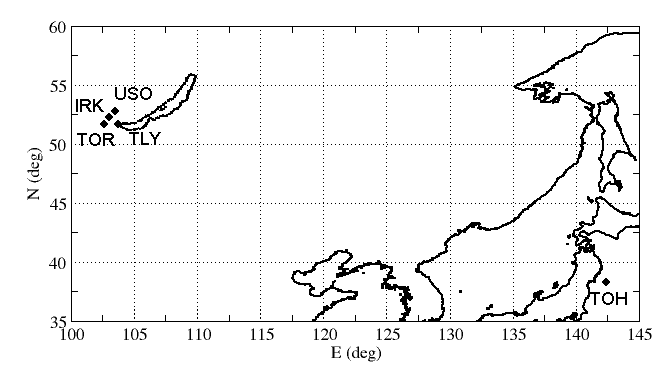} \caption{
Geometry of the observation points (IRK,TLY) and location of the earthquake epicenter(TOH) 
}
\label{fig:1} 

\end{figure}

Receiver (TOR, 51.7N, 102.6E) and transmitter (USO, 52.8N, 103.5E) of the ionosonde are located near Irkutsk (52.3N, 103E) and separated by 120km, synchronization of the time and frequency is carried out by GPS. For analysis  were used the data from TALAYA seismic station (TLY, 51.7N 103.7E), located near the ionospheric observation point. The epicenter of the earthquake was in point (TOH, 38.3N, 142.4E). Geometry of observations is shown in Fig. \ref{fig:1}. The main shock of Tohoku earthquake occurred March 11, 2011 in 05:46:24UT.

The main effect which differs the day of the earthquake from the previous and subsequent days, according to Irkutsk CHIRP ionosonde data, was the effect of midscale disturbances formation at ionograms of quasivertical sounding, manifested in the form of numerous hook-like disturbances of the main track. At Fig. \ref{fig:2}A-E there are shown examples of the experimental data during the day of the earthquake according to the Irkutsk bistatic CHIRP ionosonde. According to the study, the effect began to be observed at 06:01 UT, approximately in 14 minutes after the main shock of the earthquake.

\section{Peculiarities of plasma frequency profile}

\begin{figure}

\includegraphics[scale=0.45]{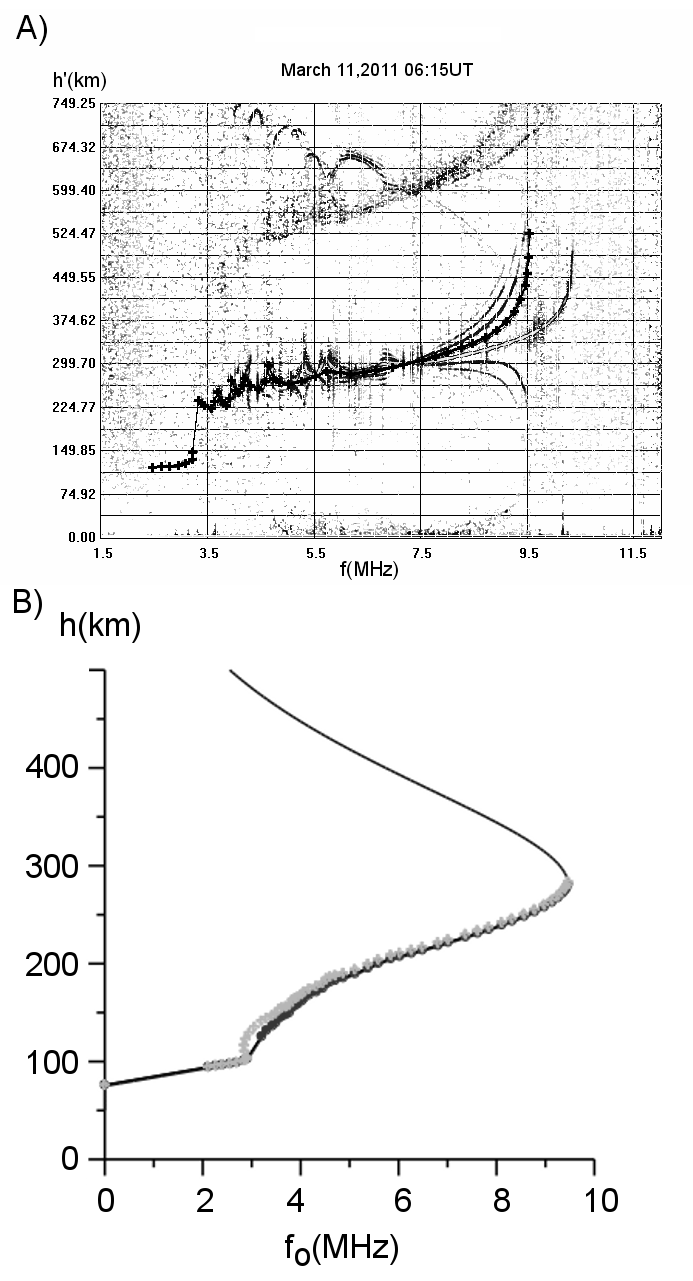} \caption{
A) Initial ionogram at 06:15UT and the result of digitizing F-layer track (marked by crosses). B) restored vertical profile of the plasma frequency in monotonic (black) and nonmonotonic (grey) assumptions. Line shows smooth profile, calculated in monotonic assumption.
}
\label{fig:3}

\end{figure}

The most significant effects of the earthquake in the ionosphere were observed at ionograms between 06:10-06:32 UT. Similar effect was seen in Japan \citep{Maruyama_et_al_2011} and Taiwan \citep{Liu_and_Sun_2011}, but at larger distances the effect apparently was not detected. For a detailed analysis and testing of analysis techniques, we chose 06:15UT  ionogram with typical distortions caused by an earthquake (Fig.\ref{fig:3}A). The line with crosses shows the result of digitizing the ordinary component of the first hop.

\begin{figure}

\includegraphics[scale=0.35]{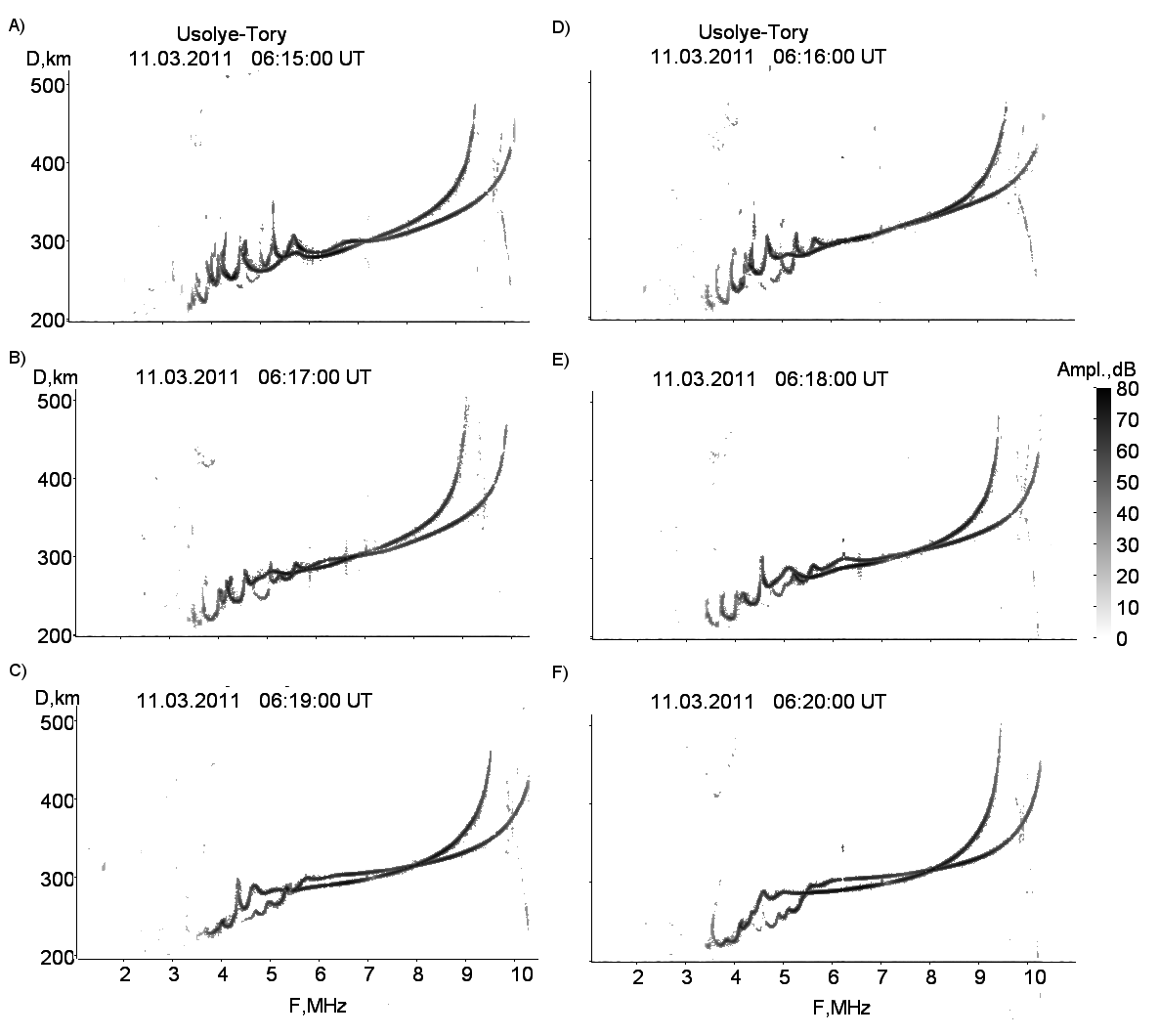}

\caption{
The appearance of the midscale ionospheric irregularities on the first hop of quasivertical ionograms during CHIRP sounding  at period 06:15-06:20UT (Irkutsk). Ionograms are filtered.
}
\label{fig:2}

\end{figure}

A typical manifestation of  the earthquake is observation of many maximums in the range-frequency characteristic (RFC) that corresponds to the reflection from the F-layer. In some cases, the track line, corresponding to the reflection from the F- layer, is even divided into multiple tracks. Each peak in the effective reflection height in the case of vertical sounding corresponds to the reflection from the bend point of the plasma frequency profile. As a consequence, from ionograms reconstruction of (Fig. \ref{fig:2} - \ref{fig:3} A) kind one should expect the emergence of multiple bend points at monotonic profile of the F-layer plasma frequency.

Important problem for calculating electron density profile from the vertical or quasivertical sounding results is the question of preserving the monotonicity of the profile while disturbances traveling during the experiment. Monotonicity is required for an unambiguous reconstruction of the profile, in the case of presense of the valleys ( nonmonotonic  areas) the solution becomes unstable. To estimate the importance of considering the profile nonmonotonicity during recovery the plasma frequency profile a reconstruction of experimental ionogram 06:15 UT was made by method of iterations \citep{MikhailovGrozov_2013} both in monotonic and non-monotonic assumptions of the plasma frequency profiles (Fig.\ref{fig:3} B). These profiles are the lower and upper boundaries for the set of profiles satisfying to the same RFC.

\begin{figure}

\includegraphics[scale=0.5]{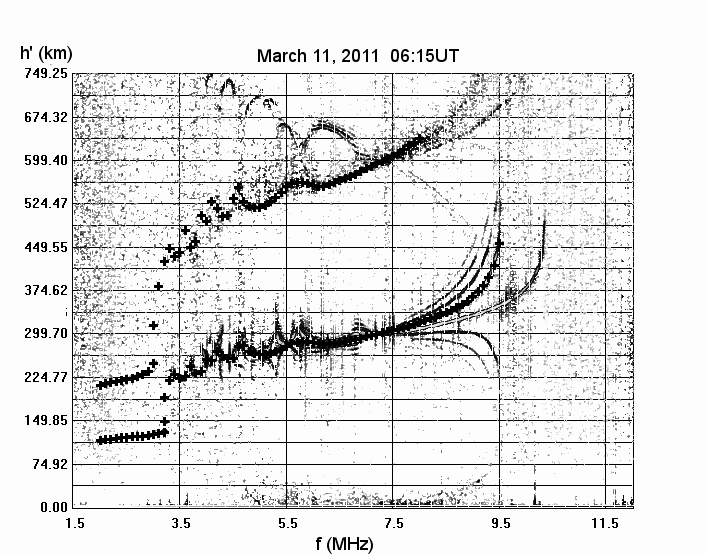}

\caption{
Ionogram of quasivertical sounding at Usolie-Tory 06:15UT. 
The crosses represent the result of the calculation of the ionogram for
reconstructed ionosphere in the monotonic profile approximation.}
\label{fig:4}

\end{figure}

Results of calculation of two-hop  RFC are obtained using monotonic profile of the plasma frequency and plotted by crosses at ionogram (Fig.\ref{fig:4}). The coincidence of the calculated and experimental RFC at single hop shows high quality of reconstruction. The coincidence of the calculated and experimental RFC for second hop proves that for short horizontal distances the ionosphere can be considered at first approximation horizontally homogeneous and the vertical irregularities can be considered as relatively largescale in the horizontal direction.

\begin{figure}

\includegraphics[scale=0.5]{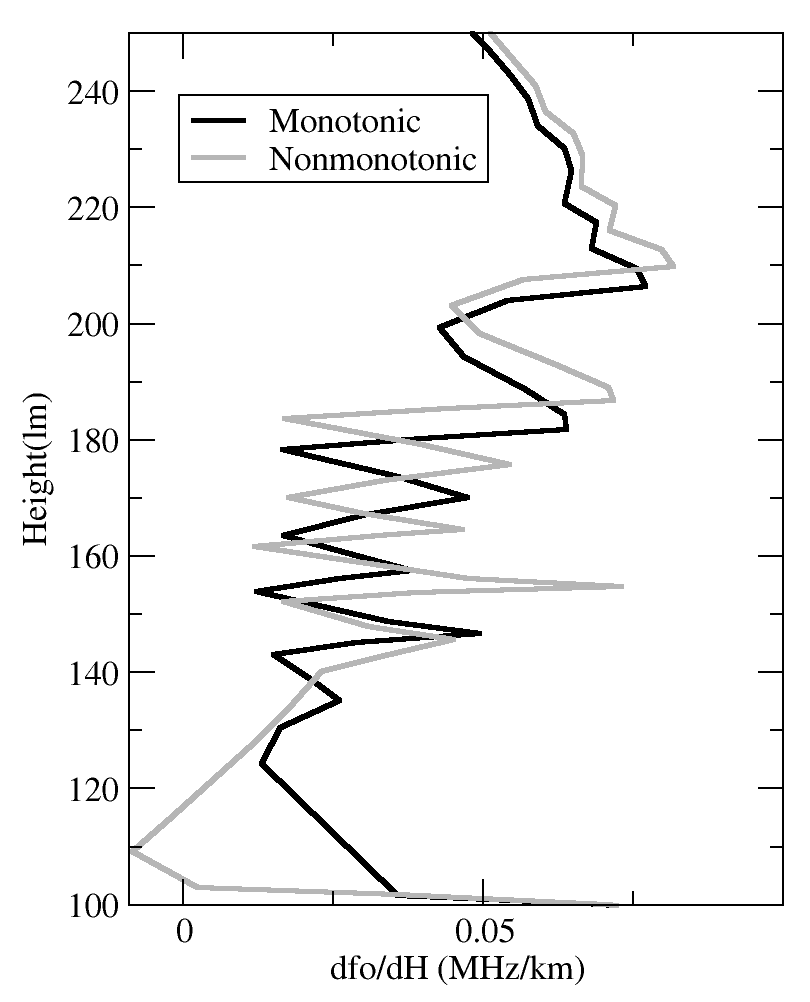}

\caption{
Altitudinal derivative of the plasma frequency, depending on the height, calculated in assumption of monotonic (black) and nonmonotonic (gray) plasma frequency profile  for the Usolie-Tory path, 06:15UT.
}
\label{fig:31}

\end{figure}

As it can be seen from Fig. \ref{fig:2}, the impact of earthquakes on the ionosphere leads to the emergence of many inflection points on the profile of the plasma frequency in the F-layer of the ionosphere. From a comparison of the plasma frequency altitude derivative for monotonic and nonmonotonic profiles(Fig.\ref{fig:31}) one can see that in assumption of nonmonotonic profile the inflection point is shifted upwards by an amount not exceeding the quasiperiod of spatial inhomogeneities. Vertical quasiperiod of inhomogeneities (defined as the distance between the zeros of the second derivative of the profile of the plasma frequency) does not change. Therefore, for calculations of the dynamics of spatial scales of vertical inhomogeneities one can use, without a significant error, inversion algorithms based on monotonicity assumption for processing experimental data, e.g. a modified Jackson's method \citep{Mikhailov2000}.

\section{Dynamics of vertical irregularities}

\begin{figure}
\includegraphics[scale=0.75]{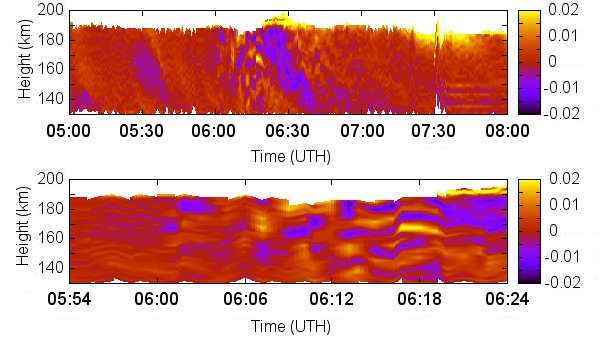} \caption{ 
Altitudinal derivative of the plasma frequency as a function of time and height. Moment of the earthquake is 05:46:24 UT.
}
\label{fig:6} 

\end{figure}

Studied in the work profiles of the plasma frequency were obtained from quasivertical ionograms with monotonic profile approximation. To select a track from ionogram we used the method described in \citep{Grozov_et_al_2012}. To highlight the time dynamics effects related to the earthquake, the variations of the plasma frequency profile have been obtained by subtracting 1-hour average profile shape. To highlight vertical structure of the irregulariteis we analized altitude variations of the frequency profile — altitude derivative of plasma frequency. The result is shown in Fig .\ref{fig:6} .

From Fig.\ref{fig:6} one can see that after the earthquake characteristic midscale irregularities appeared starting from 06:00 UT, and large-scale plasma wave appeared, negative bay of which started at 06:20UT and moved downwards. In this paper, only effects 06:00-06:20 UT will be considered.

For a detailed investigation of the effect, let us consider profile variations during the period 05:54 - 06:24 UT. The variations are shown at Fig.\ref{fig:6}. From Fig.\ref{fig:6}-B one can see several characteristic modes, the main of which begins at 05:57UT, then gradually increasing and clearly manifested at 06:01UT. From Fig.\ref{fig:6}-B one can see that the first two modes begin respectively at 06:01UT and 06:05UT, and can be analyzed as traveling up with approximate speed 100 m/s. It should be noted that the expected speed of inhomogeneties is sound speed, and in the ionosphere it increases with height, and is within 300-700m/s . Thus, the temporal resolution of the experiment (1 minute) seems to be insufficient for a detailed study of the propagation of these inhomogeneities stimulated by the earthquake.

From Fig.\ref{fig:6}-B one can see that at 06:12UT  seemingly standing waves arise with nearly zero vertical velocity. Thus, it can be assumed that the fastest modes occur due to the propagating of seismic surface wave along the earth, causing  generation of irregularities traveling upward. Slower modes are rather a mixture of the waves. It should be noted that the observed structure of oscillations at Fig.\ref{fig:6}B qualitatively correspond to the results of modeling of ionospheric effects of the earthquake, presented in \citep{Chum_et_al_2012} describing traveling of  the shock wave from the seismic vibrations upward.

~

\section{Mode composition of irregularities}

\begin{figure}

\includegraphics[scale=0.6]{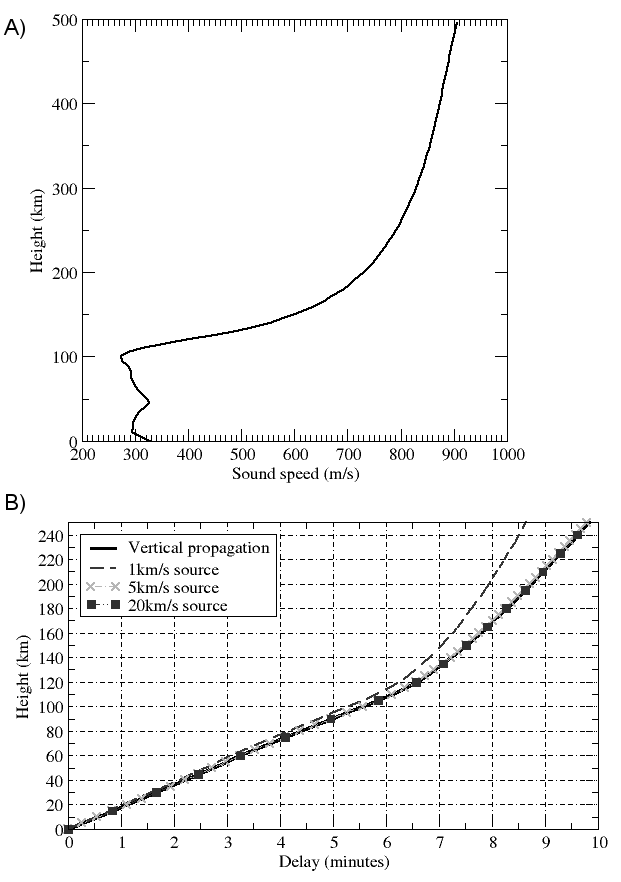} \caption{
A) The speed of sound, calculated by using NRLMSIS-00 model; B) propagation delay of the shockwave, by taking into account the Mach angle for different velocities of source (square, cross, dotted line) and in the approximation of vertical propagation of shockwave (solid line).
}
\label{fig:7}

\end{figure}

Currently, the most commonly used models is the generation of wave propagation in the ionosphere due to the passage of the shock wave cone (Mach cone) from supersonic source at the surface of the Earth \citep{Chum_et_al_2012}, and propagation of ionospheric disturbance from the center, located in the ionosphere , such as \citep{Kherani_et_al_2012}. In both cases, for calculation of the horizontal velocity of perturbation at large distances it is necessary to take into account the propagation delay of the shock disturbance from the ground to the ionosphere. To estimate this delay, by analogy with the 
\citep {Chum_et_al_2012}, we calculated propagation time for sound from the ground to the altitude of observations of main ionospheric effects (140-190km). The propagation time is from 7 to 8.5 minutes. Fig.\ref {fig:7} shows profile of the sound speed and delay profile, calculated for Irkutsk while observing the earthquake with the use of  NRLMISE-00 model.

The simulation showed that for the propagation velocities of the seismic source exceeding 1 km/s the integral error in estimating the propagation time to altitudes of 130-190km in relation to the delay calculated within approximation of vertical propagation of the shock wave is less than 60 seconds and is within the experimental accuracy. When propagating at speeds below 1km/s the error increases, and the use of the approximation of the vertical shock wave propagation produces an error that exceeds the accuracy of the measurements (60 sec.). The simulation results are shown in Fig. \ref{fig:7}. Therefore, without going beyond the limits of the experiment accuracy, for the seismic source speed faster than 1km/s the  delay for traveling of shock wave to the heights of interest can be considered as 8 minutes.

To determine the velocity-spatial structure of vertical ionospheric disturbances we made a spectral processing of altitude profiles of the plasma frequency at fixed moments

\begin{equation}
F(R,T)=\left|\int fo(h,T)e^{i2\pi h/R}dh\right|\label{eq:5}
\end{equation}

that allows to highlight the characteristic spatial scales of irregularities as a function of time. For each moment effective  horizontal velocity of disturbance has been calculated as

\begin{equation}
V=L/(T-T_{0}-dT)\label{eq:6}
\end{equation}

where $ T_{0} = 05:46:24UT $ - initial moment of the earthquake; $ dT = 00:08 $ - Propagation delay  from the ground to 140-190km heights; $ L = 3400km $ - distance along the ground from the earthquake epicenter to the observation point.

\begin{figure}

\includegraphics[scale=0.35]{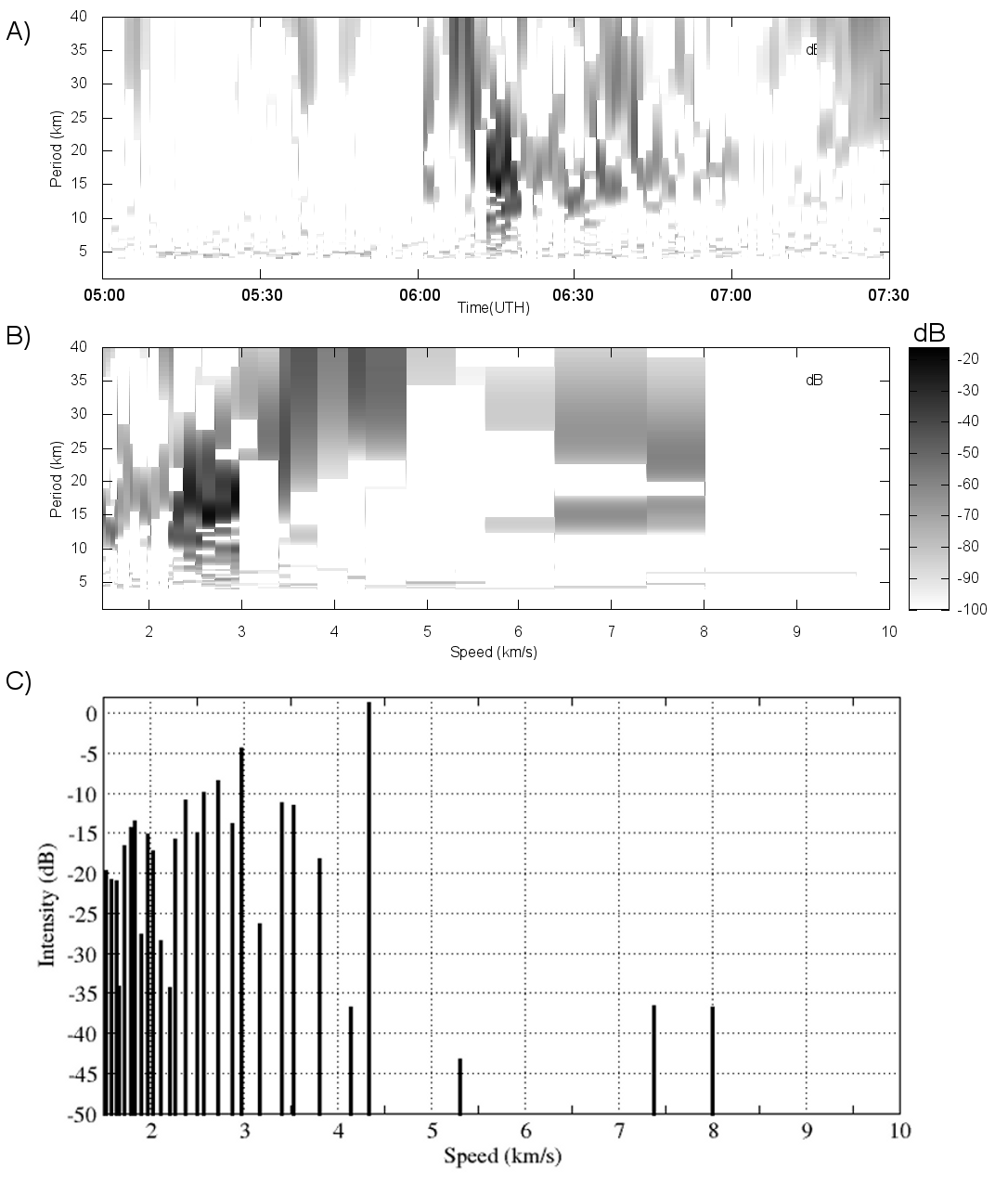} \caption{
Dependence of the quasi-period of vertical irregularities on time (A) and on the source velocity, assuming its ground movement (B). (C) - the intensity of the different spatial irregularities as function of their source velocity.
}
\label{fig:8}

\end{figure}

 Fig.\ref{fig:8}A shows the result of calculations - the function $ F(R, T) $, that determines the amplitude of vertical irregularities with specified quasiperiods as a function of the time. On Fig.\ref{fig:8}B one can see a similar function $ G (R, V) = F (R, L / V + T_ { 0 } + dT) $, that determines the amplitude of the vertical irregularities with specified quasiperiod, and calculated as a function of the horizontal velocity of irregularities travelling from the earthquake epicenter. According to the assessments above, the resulting graph $ G (R, V) $ to be within the experiment accuracy fit for the experimental estimation of horizontal velocities of propagation irregularities with speeds exceeding 1km/s. Major local maxima in distribution of the modes are summarized in Table 1.

From Fig.\ref{fig:8}B one can see that the dependence of the spatial scale on the velocity has a complex form: at slow  velocities the scale decreases with decreasing speed (30km for speed 5km/s  and 5km for the speed 2.5km/s). At velocities higher than 5km/s the spatial dependence of the period on the speed is apparently absent.

\section{Relation between the vertical structure of ionospheric irregularities and seismic fluctuations}

Let's consider the possible reasons for the formation of the mode composition of vertical ionospheric irregularities. Within the framework of geometrical optics and the model of flat Earth we assume that shock wave propagation from the seismic fluctuations in the atmosphere at different heights will be described by the expression

\begin{equation}
U(t,H)=S(t-\begin{array}{c}
H\\
\int\\
0
\end{array}cos(\alpha_{M}(r))\frac{dr}{C_{s}(r)})\label{eq:7}
\end{equation}

where $ \alpha_{M}(h) = asin \left( \frac{C_{s} (h)} {V_{seismo}} \right) $ - Mach angle; $ C_{s}(h) $ - sound velocity profile; $ V_{seismo} $ - speed of seismic wave  - the ground source of the atmospheric shock wave; $ S(t) $ - the temporal dependence of the seismic source at a given ground point.

Assuming that the speed of seismic source $ V_{seismo} $ is sufficiently high (>1km/s), the Mach angle $ \alpha_{M}(r) $ can be considered close to zero and we can consider that the shock wave is propagating almost vertically. Thus, for the vertical comparison of the structure of vertical irregularities with a shock wave we can consider function (\ref{eq:7}) in the approximation of zero Mach angle:

\begin{equation}
U(t,h)=A(h)S(t-T_{s}(h))\label{eq:9}
\end{equation}

where $ S(t) $ - time dependence of seismic vibration under investigated ionospheric point; $ A(h) $ - unimportant vertical dependence of the amplitude, which can be assumed smooth and in first approximation can be assumed as not affecting to the spatial period of irregularities;

$T_{s}(h)=\begin{array}{c}
h\\
\int\\
0
\end{array}\left[C_{s}(r)\right]^{-1}dr$
- the propagation delay of the shock front to the height h.

Function $ U (t, h) $ is a model of the acoustic signal amplitude, propagating upward from a supersonic source with temporary dependence of the amplitude $ S(t) $.
It can be shown that the spatial period of the perturbation in the shock wave increases with altitude proportional to the speed of sound. 

\begin{figure}

\includegraphics[scale=0.7]{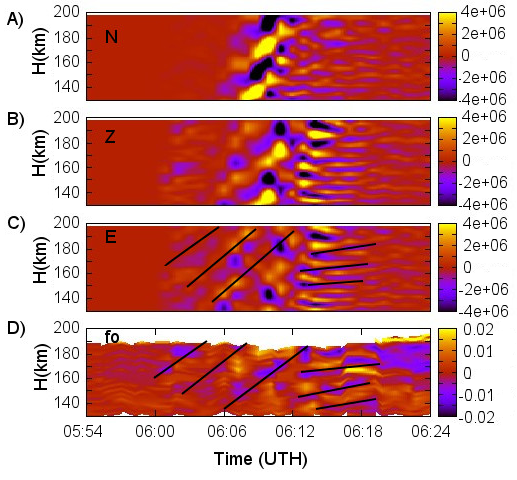} \caption{
A-C) - the altitude derivative of acoustic signal amplitude in the propagation model of the shockwave from high-speed seismic wave measured at TLY station A) - the north-south component, B) - vertical component C) - the west-east component). These models of the neutral irregularities are smoothed for imitation of the experimental detection. D) The altitude derivative of the plasma frequency is obtained from the experimental data of the Irkutsk CHIRP ionosonde
}
\label{fig:11}

\end{figure}

For the simulation, as time dependence of the amplitude $ S(t) $ we used data from seismic observations at TLY observatory, located near the ionospheric observations point. The simulation results are shown in Fig.\ref{fig:11}. 
The figure shows that the expected neutral irregularities associated with the passage of shock cone waves have a finer spatial and temporal structure than it is observed in the experiment. This shows that a detailed diagnosis of the propagation of ionospheric irregularities which arise due to propagation of the shock wave from the seismic vibrations can not be done because of insufficient temporal resolution at Irkutsk CHIRP ionosond.

For comparison of simulation results with experimental data there was made an imitation process for conducting ionospheric measurements. For imitation of sounding process and processing the sounding data the  model (\ref{eq:9}) were decimated to obtain 60 seconds time resolution and smoothed by moving 10km window over the height (simulating the possible altitude shift due to probable nonmonotonic profiles and simulating the effects of spatial averaging during sounding and processing). The results of such processing of  the model data (\ref{eq:9}) and their comparison with the ionospheric irregularities are shown at Fig.\ref{fig:11}.

The Fig.\ref{fig:11} shows that the structure of the averaged and decimated data differs significantly from the structure of irregularities formed by the shock wave . The Fig.\ref{fig:11} shows that the structure of ionospheric disturbances during 06:00-06:12UT is qualitatively similar to the decimated and filtered structure of the shock wave, and the most likely the sources of the observed ionospheric irregularities are Z or/and E component of seismic vibrations. This can be explained by the geometry of the experiment ( Fig.\ref{fig:1} ) - seismic wave propagates almost from the east. The impact of E- component of seismic disturbance to the ionosphere can also be explained by complex mountainous terrain of the Baikal region. From Fig.\ref{fig:11}A,D it is seen that simulated atmospheric effect from  N- component of seismic oscillations starts much later than the beginning of ionospheric variations, so , apparently, N-component is not the main cause of them.

In all cases (Fig.\ref{fig:11} A-D) during the period 06:00-06:12UT one can observe monotonous upward movement of irregularities associated with the propagation of the acoustic shock wave. During 06:12-06:20UT the acoustic signal according to the simulation and the experimental data looses its apparent structure  of upward propagating waves, and becomes having apparent structure of stationary waves, reflecting the features of spatial interference of acoustic oscillations and their period. Both of these effects reflect the structure of the acoustic signal in the shock wave produced by the model (\ref{eq:9}). This allows to conclude that from 06:00 to 06:20UT the main ionospheric effects are associated with the traveling of a shock wave cone from a ground seismic source, and after 06:20UT the ionospheric effects are apparently associated with not a supersonic ground source, but with a different mechanism of disturbance generation.

Similar effects of the ionospheric repetition of seismic vibrations of the Earth have been observed in particular during  Doppler sounding of the ionosphere during the Tohoku earthquake \citep{Chum_et_al_2012}, as well as in the bottom of the ionosphere while creating artificial seismic disturbances by seismic vibrators \citep{Kuznetsov_et_al_1999}. The results allow to verify different models, for example \citep{Kherani_et_al_2012, Maruyama_and_Shinagawa_2014}.

~

\section{Conclusion}

In the paper we present results of observations of midscale vertical ionospheric irregularities caused by the Tohoku earthquake 11/03/2011 according to Irkutsk CHIRP ionosonde at 3400km distance from the epicenter. Relationship of these irregularities with the earthquake can be substantiated by observations of the similar irregularities after the earthquake in relative proximity to the epicenter \citep{Maruyama_et_al_2011, Liu_and_Sun_2011}.

Experimental data allow us to suggest the observation of several modes of spatial oscillations, presumably propagating with different velocities (Fig.\ref{fig:6}). With taking into account the vertical velocity upward propagating acoustic waves, by analogy with \citep {Chum_et_al_2012} there was designed a spatial spectrum of irregularities, as a function of the horizontal velocity in the range of 1.5-10km/s (Fig.\ref{fig:8}B) and as a function of time (Fig.\ref{fig:8}A). Parameters of the mode composition of the irregularities are summarized in Table 1.

We simulated acoustic shockwave propagation caused by seismic vibrations (based on seismic data from TLY observatory, located near the site of ionospheric observations) considering the effects of sound propagation velocity, which depends on altitude. The simulation showed that this model, averaged by taking into account the peculiarities of the experiment sufficiently well describes the characteristics of the modes of vertical ionospheric irregularities during the period 06:00-06:20 UT. During 06:00-06:12UT it predicts apparent upward traveling wave packets, and during 06:12-06:20 – stationary waves are apparent (Fig.\ref{fig:11}) 

The most likely sources of these ionospheric irregularities were Z and E -components of seismic vibrations. Similar relationship have been observed, in particular, in Doppler sounding of the ionosphere during this earthquake \citep{Chum_et_al_2012}, and in the lower part of the ionosphere when creating artificial ionospheric disturbances by seismic vibrators \citep{Kuznetsov_et_al_1999}. The results allow to verify different models, for example \citep{Kherani_et_al_2012, Maruyama_and_Shinagawa_2014}.

\section{Acknowledgements}

This work was supported by the DPS RAS Program project IV.12.2 (registration number 01201255931). The authors are grateful to  Global Seismograph Network  (GSN-IRIS/IDA) for the use of seismic observatory TLY data. The authors are grateful to Perevalova N.P. (ISTP SB RAS) for fruitful discussion.

~


\newpage{}

\newpage{}

\newpage{}

\newpage{}

\newpage{}

\newpage{}

\newpage{}

\newpage{}

\newpage{}
\begin{figure}

\includegraphics[scale=0.7]{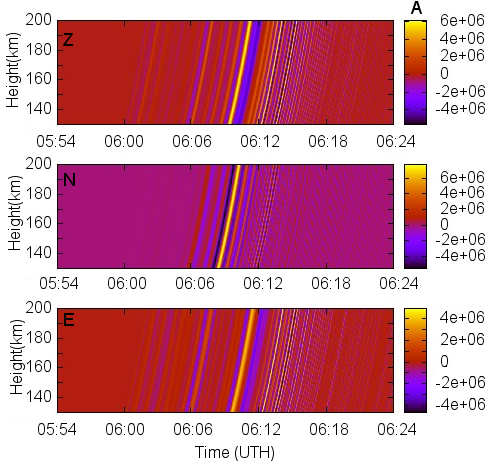} \caption{
Simulation of acoustic irregularities generated by the shock wave from the seismic source, defined from TLY seismic data.
}
\label{fig:10}

\end{figure}

\newpage{}

\newpage{}
\begin{table}
\caption{ Characteristics of individual motion modes of ionospheric waves according to the vertical structure of the irregularities, measured in Irkutsk.
}

\begin{tabular}{|c|c|c|c|}
\hline 
No  & Period(km)  & Velocity(km/s)  & Amplitude(dB)\tabularnewline
\hline 
\hline 
1  & 20-40  & 6.5-8  & -35\tabularnewline
\hline 
2  & 15  & 6.5-8  & -35\tabularnewline
\hline 
3  & 25-40  & 4.2-5  & 0\tabularnewline
\hline 
4  & 15-40  & 3.2-4  & -10\tabularnewline
\hline 
5  & 12-25  & 2.3-3  & -5\tabularnewline
\hline 
6  & 10-40  & 1.5-2.2  & -15\tabularnewline
\hline 
\end{tabular}
\end{table}

~

~
\end{document}